\documentclass[conference,a4,comsoc]{IEEEtran}
\usepackage{graphicx}
\usepackage{amssymb}
\usepackage{amsmath}
\usepackage{cite}
\usepackage{mathrsfs}
\usepackage[displaymath,mathlines]{lineno}
\usepackage{color}
\usepackage{tabulary}
\usepackage{multirow}
\usepackage{algpseudocode}
\usepackage{pbox}
\usepackage{multicol}
\usepackage{lipsum}
\usepackage{algorithm}
\usepackage{algpseudocode}
\usepackage{amsthm}
\usepackage{relsize}
\usepackage{lipsum}
\usepackage{epstopdf}
\usepackage{times}
\usepackage{enumitem}
\usepackage{float}
\usepackage{mathtools}
\usepackage{calc}
\usepackage{makecell}
\usepackage{caption}
\usepackage{subcaption}
\usepackage{soul}
\usepackage{algorithm}
\usepackage{fancyhdr}
\usepackage{verbatim}
\makeatletter
\newcommand{\doublewidetilde}[1]{{%
		\mathpalette\double@widetilde{#1}}}
\newcommand{\double@widetilde}[2]{%
		\sbox\z@{$\m@th#1\widetilde{#2}$}%
		\ht\z@=.5\ht\z@
		\widetilde{\box\z@}}
\makeatother

\allowdisplaybreaks

\makeatother
\makeatletter

\usepackage{titlesec}

\begin{document}

\title{\huge Clustering Strategies in Satellite-Aided Communications \vspace{-0.25cm}
}
\author{Tam Ninh Thi-Thanh$^{*,\xi}$, Nguyen Minh Quan$^\xi$, Do Son Tung$^\xi$, Trinh Van Chien$^\xi$, Hung Tran$^\nu$
\\\fontsize{9}{9}{\textit{$^*$Faculty of Information and Communication Technology, National Academy of Education Management, Vietnam}}
\\ \fontsize{9}{9}{\textit{$^\xi$School of Information and Communications Technology, Hanoi University of Science and Technology, Vietnam}}
\\\fontsize{9}{9}{\textit{$^\nu$DATCOM Lab, Faculty of Data Science and Artificial Intelligence, National Economics University, Vietnam}} 
\\\fontsize{9}{9}{\textit{Corresponding author:} Nguyen Minh Quan}%
\vspace{-0.5cm}
}
\maketitle


\begin{abstract}

With the rapid advancement of next-generation satellite networks, addressing clustering tasks, user grouping, and efficient link management has become increasingly critical to optimize network performance and reduce interference. In this paper, we provide a comprehensive overview of modern clustering approaches based on machine learning and heuristic algorithms. The experimental results indicate that improved machine learning techniques and graph theory-based methods deliver significantly better performance and scalability than conventional clustering methods, such as the pure clustering algorithm examined in previous research. These advantages are especially evident in large-scale satellite network scenarios. Furthermore, the paper outlines potential research directions and discusses integrated, multi-dimensional solutions to enhance adaptability and efficiency in future satellite communication.

\end{abstract}

\section{Introduction}

Satellite communications networks (SATCOM) are considered a promising solution to meet the connectivity demands of a fully connected world, especially in the 6G era \cite{guidotti2020architectures}. There are three main categories of satellites characterized by their orbital altitudes, specifically the Geostationary Earth Orbit (GEO), Medium Earth Orbit (MEO), and Low Earth Orbit (LEO). While GEO satellites support reliable connectivity and service continuity for terrestrial \cite{jung2024modeling}, non-geostationary orbit (Non-GSO) satellites like MEO and LEO have the potential to provide low latency, improve communication speeds, and allow many more applications that require high quality of service (QoS). In recent years, with advances technologies and techniques like beamforming, beam hopping, dynamic spectrum access, and intelligent resource management, mega-constellations LEO satellites have been proposed to allow users in remote areas to use Internet services, where terrestrial coverage is struggle to reach \cite{10507224, 10713888, park2025trends}. 
 
Nonetheless, satellite communication networks face persistent technical challenges that impact system efficiency and scalability. One such challenge is the scheduling of satellite-ground communications for telemetry, tracking, and command tasks across multiple satellites and ground stations \cite{song2023cluster}. Another important issue is efficient beam placement, which is essential for maximizing spectral efficiency, improving link budgets, and balancing the traffic load across beams in ultra-dense LEO constellations \cite{10798457}. Beam hopping, while offering flexibility, introduces additional complexity in practice, including real-time beam scheduling, adaptive power allocation, and coordination between moving satellites and diverse ground user demands, especially under fast-changing link conditions \cite{xie2025multi}. Furthermore, rapid fluctuation in spectrum availability can produce unpredictable interference patterns, further complicating resource management and necessitating robust interference mitigation strategies\cite{10946453}. The determination of the SDN controller in a topology that changes rapidly and has limited inter‐satellite bandwidth is also considered a challenge to preserve network performance \cite{bukhari2025k}. To address these challenges, clustering strategies have emerged as one of the practical solutions to manage the increasing complexity and scale of satellite communication networks \cite{jung2023satellite}. By grouping users or satellites according to geographic location, data traffic, or connectivity, clustering helps optimize resource allocation, interference reduction, and reduces communication overhead \cite{liu2021reliable}. Furthermore, the use of machine learning and heuristic algorithms in the clustering process enables the system to adapt in real time to changing link conditions and user demands, offering a flexible and efficient solution for future high-density satellite networks.

Despite growing interest in clustering strategies for satellite communication networks, existing research remains fragmented. Many studies focus on specific scenarios or rely on case-dependent assumptions, limiting the generalizability of their findings across diverse satellite configurations and service requirements. As a result, there is a clear gap in establishing a unified framework for systematically evaluating and comparing clustering techniques under the dynamic and heterogeneous conditions of satellite networks. Motivated by this, our objective is to provide a comprehensive review of recent advances in clustering algorithms for satellite communications. We aim to bridge the gap between theoretical models and practical implementations by highlighting key challenges, synthesizing current methodologies, and identifying open research directions. This work serves as a foundation for designing scalable, adaptive, and intelligent clustering frameworks tailored to next-generation satellite networks.

The main contributions of this work are summarized as follows: $i)$ we conduct a comprehensive review and comparison of clustering techniques, ranging from classical machine learning algorithms such as $K$-Means clustering to heuristic approaches. We evaluate their computational efficiency, scalability, and responsiveness to the dynamic, delay-sensitive, and resource-constrained nature of satellite communication networks; $ii)$ we emphasize the rise of hybrid frameworks that integrate conventional optimization tools with data-driven learning models. These approaches are shown to effectively address non-convex, large-scale clustering problems under uncertain and time-varying satellite channel conditions; and $iii)$ we consolidate recent research trends and identify fundamental open problems to propose key research directions to guide future work toward more robust and scalable  networks.

\begin{figure}
    \centering
    \includegraphics[trim={0.5cm 3cm 0.1cm 3cm},clip,width=0.8\linewidth]{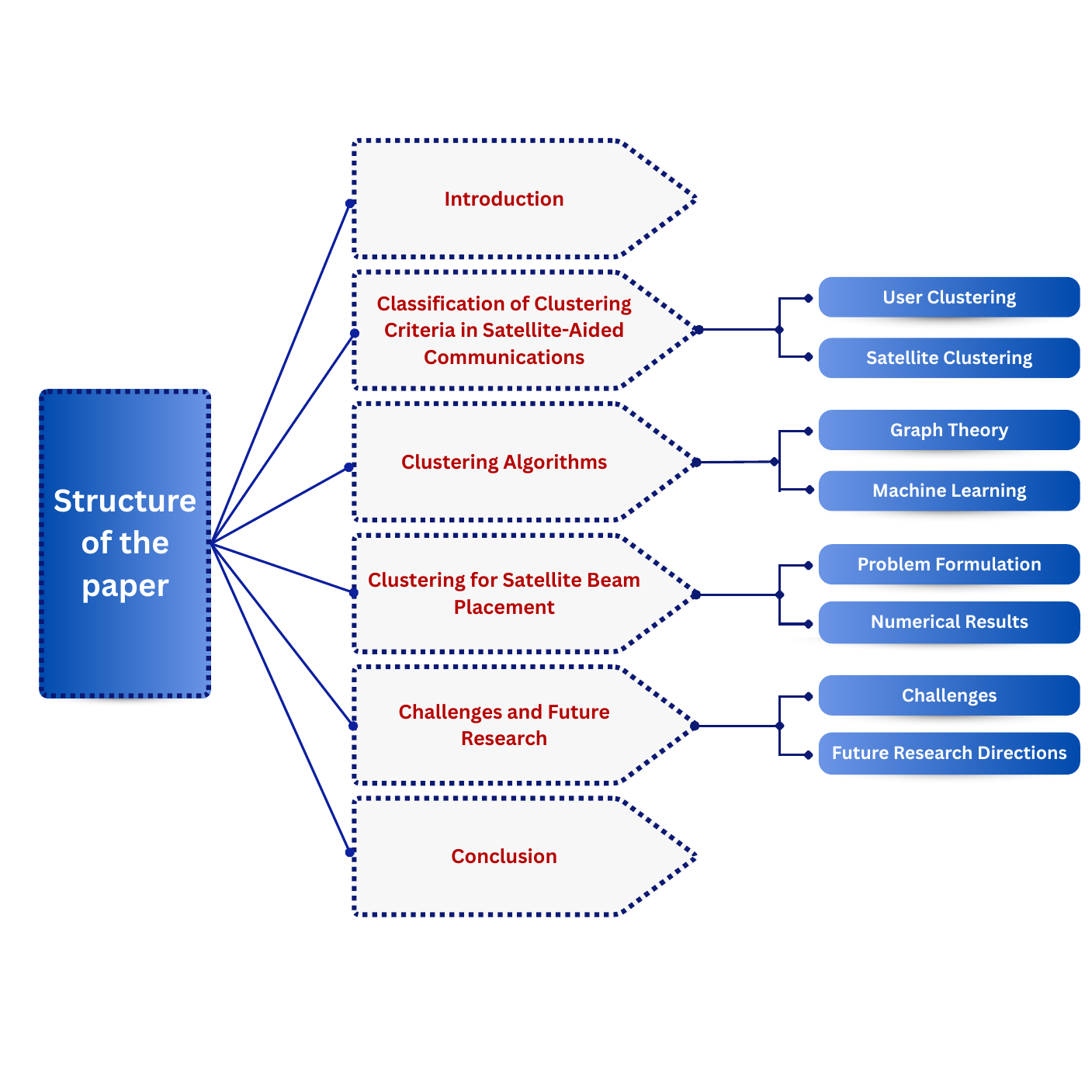}
    \caption{Structure of Machine Learning and Multi-Approach Clustering Strategies in Satellite Communication Systems}
    \label{fig:structure}
    \vspace{-0.5cm}
\end{figure}

\section{Classification of Clustering Criteria in Satellite-aided Communications}

\subsection{User Clustering}

In modern satellite communication systems, the growing number of user terminals places significant pressure on limited resources such as bandwidth, power, and beam capacity~\cite{9295418}. Without effective allocation, high user density can lead to performance degradation 
while also increasing the computational complexity of resource scheduling and coordination, challenging scalability, and real-time operation. User clustering offers a practical approach to these challenges by organizing users with similar signal characteristics, geographical positions, or service demands. This reduces system complexity by shifting resource allocation from the individual to the group level, lowering computational overhead, improving overall efficiency and capacity \cite{10445284}. We now introduce three primary user clustering criteria as follows

\textit{$1)$ Geographical Location}: One of the primary criteria for clustering users in satellite communication systems is their geographical location. Satellite beams are often designed to cover specific areas, and user clustering based on geographical locations ensures efficient allocation of beams. By grouping users located within geographical regions with similar conditions such as propagation environment, mobility patterns, interference levels, and weather effects, the network can reduce the overhead associated with handover and inter-beam interference. This enables more effective resource management and ensures seamless communication ~\cite{10648635,guidotti2020clustering}.

\textit{$2)$ Channel Characteristics}: Another approach to clustering users is based on similarities in their channel characteristics, such as channel vectors, path loss, and signal-to-noise ratios \cite{10034945}. This approach leverages the physical layer properties of the communication channel, enabling the system to group users with correlated or orthogonal channel state information (CSI). By clustering users with similar channel conditions, it becomes possible to apply more efficient precoding strategies, reduce inter-user interference, and improve spectral efficiency. In addition, it could improve both channel capacity and bit error rate (BER), particularly in high SNR conditions \cite{10433460}. This is especially relevant for LEO multibeam systems employing co-frequency reuse and OFDM, where inter-beam interference is more severe than in traditional GEO systems.

\textit{$3)$ Service Demand}: 
Service demands are another important factor when clustering users in satellite networks \cite{10601666}. Different users may require varying levels of service quality based on their application needs, such as voice calls, video streaming, or data downloads. By clustering users according to their service demands, we can optimize bandwidth allocation to meet the specific needs of each group with wireless multicast technology \cite{10034945}. For example, users with high bandwidth demands can be grouped to ensure that sufficient resources are available to maintain the required throughput, while users with lower demands can share bandwidth more efficiently \cite{10460317}.
\subsection{Satellite Clustering}
While user clustering focuses on grouping users to enhance service efficiency,
satellite clustering shifts focus toward the structural and functional organization of satellites themselves. This aspect becomes particularly critical in LEO constellations, where hundreds or thousands of satellites must cooperate under dynamic topologies and limited inter-satellite links.
By forming clusters of satellites, the overall system can reduce the complexity of overhead management, respond to real-time communication demands, and maintain robust performance in a dynamic environment. We classify it into three principal criteria as follows

\textit{$1)$ Communication Objectives}: To support global coverage with high-speed and low-cost connectivity, current satellite networks commonly adopt a multi-layer architecture that combines a large number of LEO satellites with a smaller number of GEO and MEO nodes. While this structure improves coverage and flexibility, it also increases the complexity of network management. To address this, recent work has explored clustering LEO satellites based on predicted service demands instead of relying solely on topological configuration. Such service-aware clustering has been shown to improve service  and enhance overall system performance \cite{zhu2022service}.

\textit{$2)$ Functional Roles:} 
For large-scale satellite constellations with thousands of nodes, effective network management and adaptability to dynamic conditions are critical, yet traditional decentralized methods become inefficient and hard to scale. Clustering addresses these issues by logically grouping satellites, which reduces management overhead and enhances scalability and maintainability~\cite{liu2021reliable}. In TT\&C (Telemetry, Tracking, and Command) applications, the increasing number of LEO satellites poses challenges for real-time processing and selection. Due to hardware and computational limitations, receivers cannot process all visible satellites simultaneously. By clustering satellites with similar geometric characteristics, the system can select a smaller, representative subset, reducing computational load and improving positioning accuracy while adapting to the fast-changing geometry of LEO constellations~\cite{10648888,jing2025fast}.

\begin{figure}[t]
	\centering
    \begin{minipage}[t]{0.49\textwidth}
        \centering
        \includegraphics[trim=3.0cm 3.0cm 3.0cm 3.6cm, clip=true, scale=0.35]{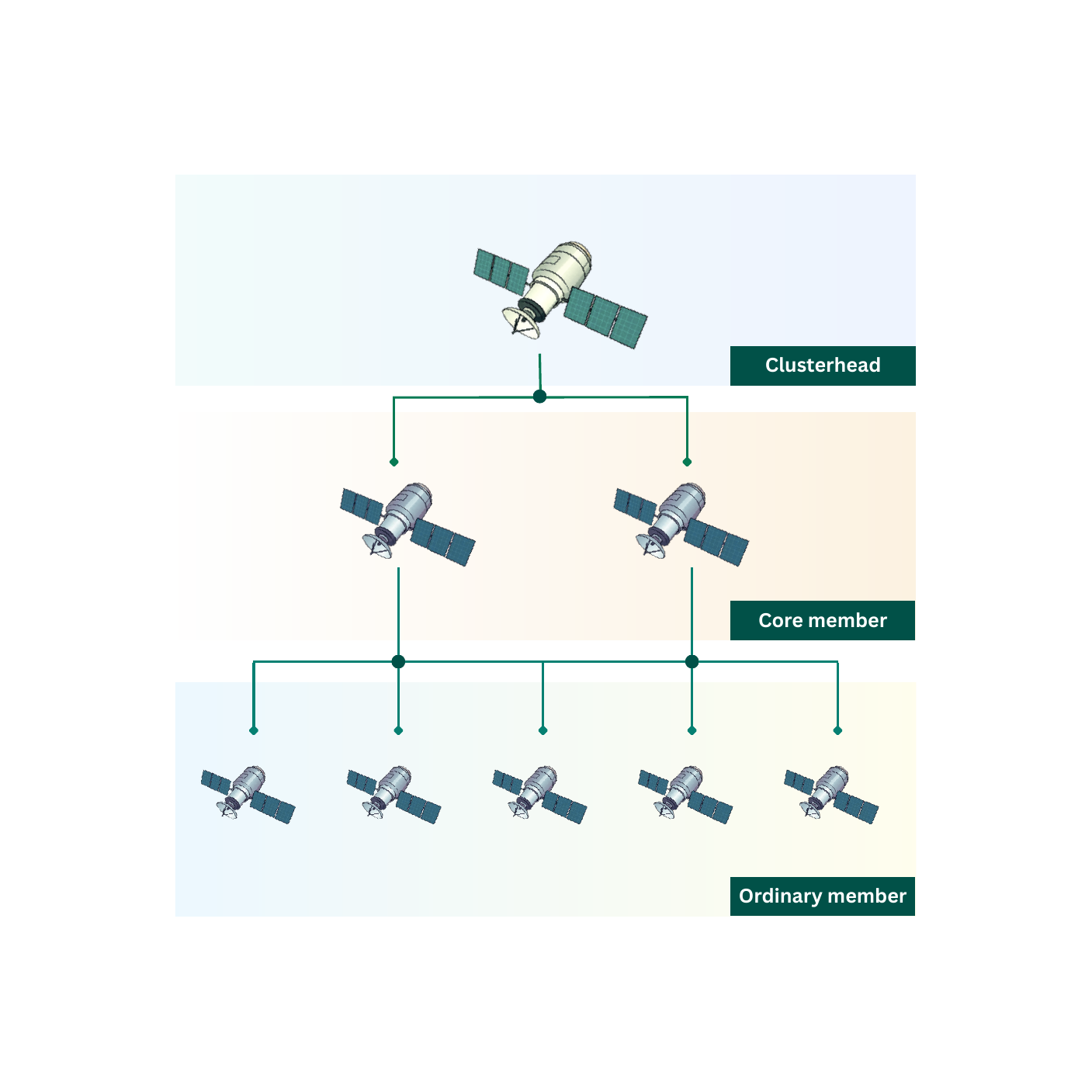}    	
        \subcaption{Hierachical clustering}
    \end{minipage}
    \hfill
    \begin{minipage}[t]{0.49\textwidth}
        \centering
        \includegraphics[trim=0.0cm 0.0cm 0.0cm 2cm, clip=true, scale=0.32]{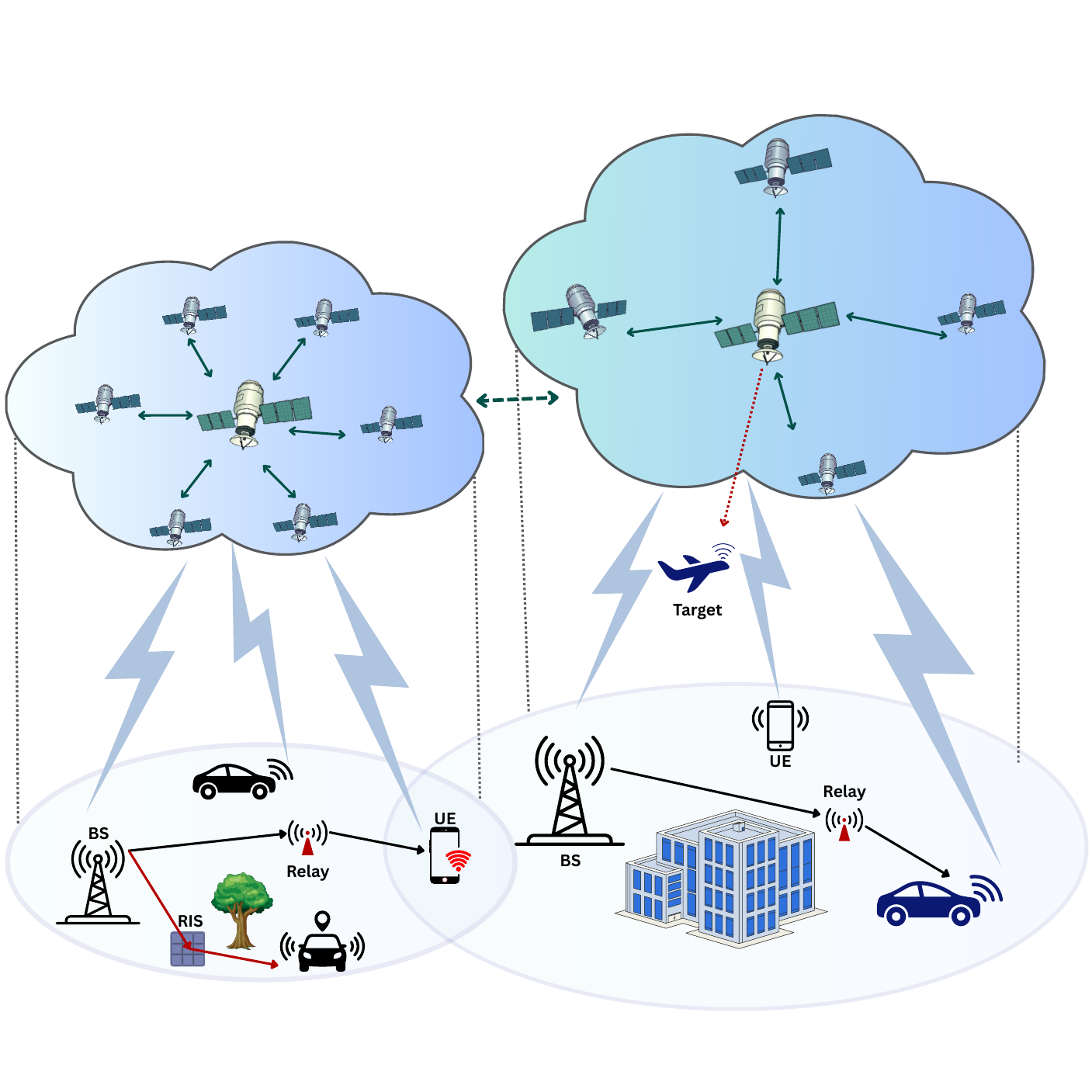}
        \subcaption{Clustering Satellite}
    \end{minipage}
\vspace{-0.2cm}
\caption{The clustering stratergies in satellite-aided communications: $(a)$ Hierarchical clustering; and $(b)$ Clustered satellites integrated into mobile networks.}
\label{fig:figcluster}
\vspace{-0.5cm}
\end{figure}

\textit{$3)$ Network Topology}: A well-designed cluster architecture not only improves the reliability of satellite operations but also facilitates robust fault tolerance mechanisms, ensuring high reliability, availability, and QoS \cite{geng2021resilient}. Fig.\ref{fig:figcluster}(a) illustrates a hierarchical satellite clustering structure used in a resilient communication framework to handle uncertain network threats. In this structure, satellites are grouped into smaller clusters, each managed by a cluster head, which enhances stable control, efficient data fusion, and fault tolerance. 

Fig.~\ref{fig:figcluster}(b) illustrates a satellite cluster architecture based on the concept in~\cite{jung2023satellite}, where functions are divided between a master and multiple slave satellites. This functional separation reduces individual payloads, simplifies satellite design, and improves scalability.
In the context of 6G networks, with increasing integration among terrestrial, aerial, and satellite segments, such cluster-based architectures are essential to enable flexible, resilient, and intelligent communications. Satellites form cooperative clusters that facilitate efficient data exchange and coordinated control between distributed nodes. These clusters interact with terrestrial base stations (BS), relay nodes, reconfigurable intelligent surfaces (RIS), and user equipment (UE), creating a highly interconnected multi-layer network. This architecture supports dynamic task offloading, adaptive resource allocation, and service continuity across diverse environments, from dense urban zones to highly mobile scenarios like vehicular networks \cite{tung2024graph}.



\section{Clustering Algorithms}

\subsection{Graph Theory}

Graph theory plays a crucial role in addressing clustering problems within satellite communication, particularly for dynamic and large-scale networks such as LEO constellations. By modeling agents (i.e., satellites or users) and their relationships as nodes and edges in a graph, clustering algorithms can efficiently group these agents, which are essential for optimizing routing, interference management, and resource allocation. We can define an undirected graph $\mathbf{G} = (V, E)$, where $V$ is a set of vertices representing the agents, and $E = {(u_1, u_2), u_1, u_2 \in V}$ is a set of edges, where each edge $(u_1, u_2)$ indicates the existence of a communication or similarity relationship between two agents. This graph can be represented by an adjacency matrix $\mathbf{H}$ defined as:
$\mathbf{h}_{u_1u_2}$ is 1 if a connection or relationship exists, and is 0 otherwise. Given that no agent is connected to itself, the adjacency matrix $\mathbf{H}$ is symmetric with all diagonal entries equal to zero, and composed of nonnegative values. In~\cite{10018884}, a graph-based flocking control framework was proposed using artificial potential fields and a degree-constrained spanning tree to achieve collision-free clustering in satellite formations. Using Prim’s and Greedy algorithms, the framework achieves limited connectivity, low delay, and stable formations with low complexity.
In~\cite{9914723}, a greedy graph-based clustering strategy was introduced for user scheduling in LEO-MIMO systems. Using the MaxCliqueDyn algorithm, users with uncorrelated channels are grouped efficiently, reducing inter-beam interference and improving system capacity. In~\cite{10798457}, a greedy graph coloring algorithm was developed to minimize the number of active beams while maintaining acceptable user gain under geographical constraints. This low-complexity method is well-suited for dynamic environments.

\subsection{Machine Learning}

Machine learning (ML) refers to a class of data-driven algorithms that identify patterns and make decisions without explicit programming. These techniques have proven effective in solving complex optimization problems, especially in dynamic and uncertain environments. This makes ML well-suited for satellite systems, which operate over non-deterministic channels, have time-varying topologies, and face strict resource constraints. Unsupervised learning algorithms are particularly valuable for tasks like clustering and user grouping, as they can uncover hidden structures in unlabeled data. Common examples include K-Means Clustering, self-organizing maps (SOM), hidden Markov models (HMM), and restricted Boltzmann machines (RBM), each offering strengths in pattern recognition and data aggregation~\cite{5453745}.

\textit{$1)$ $K$-Means Clustering} is one of the most widely used algorithms for clustering due to its simplicity and computational efficiency. It aims to partition a dataset into $K$ disjoint clusters such that data points within the same cluster are as close as possible to each other, while data points in different clusters are as far apart as possible. The algorithm proceeds iteratively and consists of the following principal steps:

\begin{enumerate}
  \item \textbf{Initialization:} Randomly initialize $K$ centroids $\mu_1, \dots, \mu_K$.
  \item \textbf{Assignment:} Assign each point $x_i$ to the nearest cluster $C_k$ based on Euclidean distance: $k = \arg\min_j \|x_i - \mu_j\|^2.$

  \item \textbf{Update:} Recompute the centroid of each cluster: $\mu_k = \frac{1}{|C_k|} \sum_{x_i \in C_k} x_i.$
  \item \textbf{Repeat:} Repeat steps 2–3 until convergence (i.e., centroids do not change or an error threshold is reached).
\end{enumerate}



In satellite communications, K-Means has been effectively applied to various clustering problems. In~\cite{10133823}, a modified K-Means algorithm was used to reduce the number of active beams and improve load balance with low complexity. Yuan \textit{et al.}~\cite{10794213} employed K-Means++ to cluster users based on geographical location and traffic demand, enabling adaptive beam layouts that enhance system efficiency, coverage, and resource utilization. In~\cite{camino2021milp}, K-Means was integrated into a MILP framework to optimize irregular beam layouts by grouping users with similar locations and traffic patterns. These clusters served as spatial constraints for beam placement, improving scalability and computation speed while maximizing covered traffic.


\textit{$2)$ Advanced Learning-Based Clustering}:
The authors in \cite{chen2023edge} propose a federated learning framework to reduce transmission latency and overhead. It integrates a LEO Edge Selection and Clustering (LESC) mechanism to cluster LEO satellites based on SNR thresholds, ensuring high-quality communication during model updates. A re-clustering mechanism is also introduced to adapt to LEO mobility, maintaining efficient and accurate training.

\begin{figure*}[ht]
	\centering
    \begin{minipage}[t]{0.245\textwidth}
        \includegraphics[trim=4.3cm 8.0cm 4.7cm 9cm, clip=true, scale=0.35]{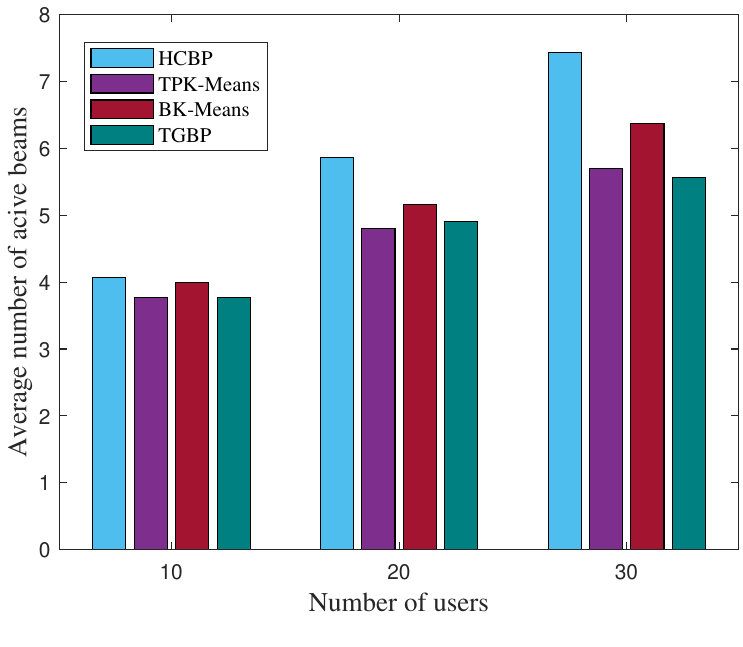}  
        \subcaption{}  	
        \label{fig:number_of_beams_small}
    \end{minipage}
    \begin{minipage}[t]{0.245\textwidth}
	\includegraphics[trim=4.3cm 8.0cm 4.7cm 9cm, clip=true, scale=0.35]{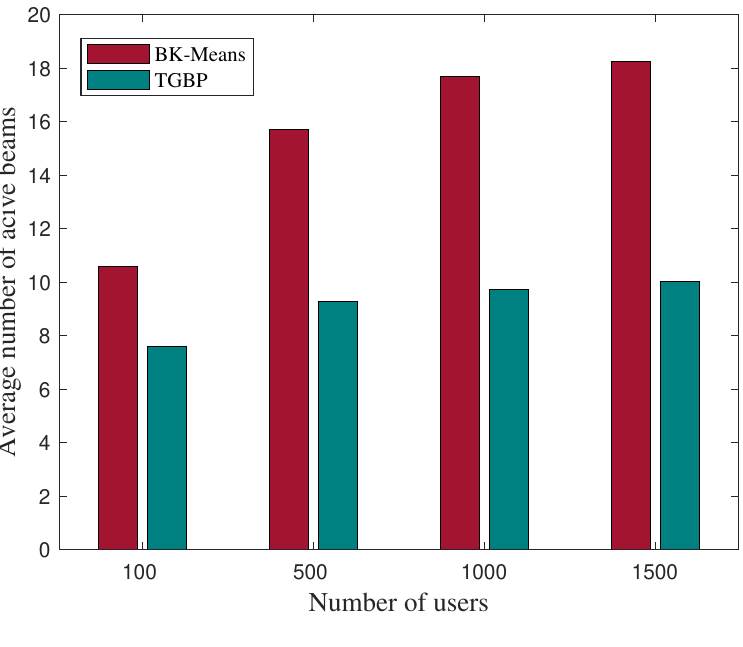}
        \subcaption{}
        \label{fig:number_of_beams_large}
    \end{minipage}
    \begin{minipage}[t]{0.245\textwidth}
	\includegraphics[trim=4.3cm 8.0cm 4.7cm 9cm, clip=true, scale=0.35]{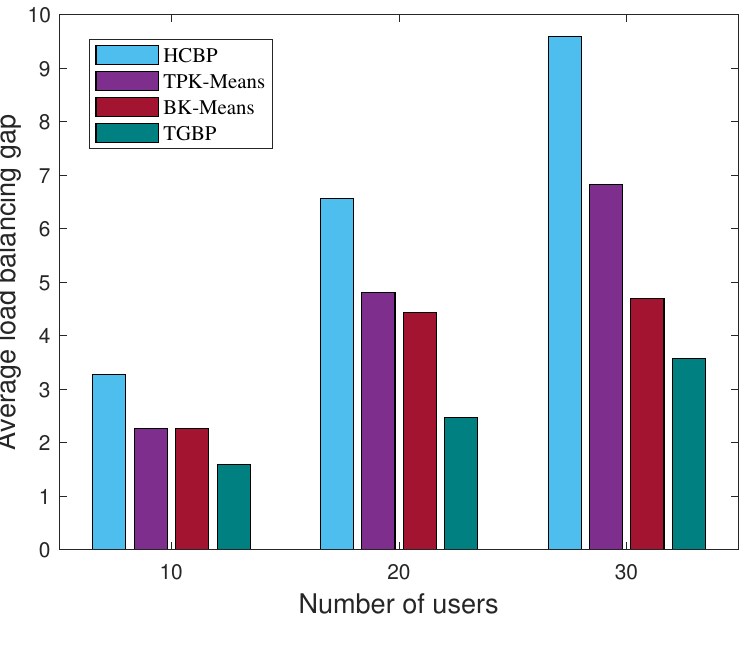}
        \subcaption{}
        \label{fig:balancing_gap_small}
    \end{minipage}
     \begin{minipage}[t]{0.245\textwidth}
	\includegraphics[trim=4.3cm 8.0cm 4.7cm 9cm, clip=true, scale=0.35]{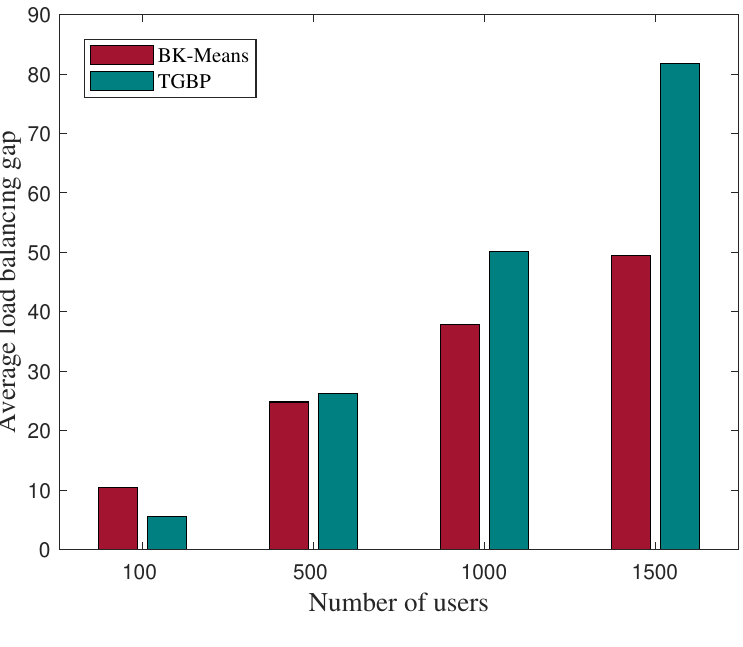}
        \subcaption{}
        \label{fig:balancing_gap_large}
    \end{minipage}
\vspace{-0.2cm}
\caption{(a) Number of active beams on the small user set. (b) Number of active beams on the large user set. (c) Average load balancing gap on the small user set. (d) Average load balancing gap on the large user set. }
\label{fig:fig2}
\vspace{-0.3cm}
\end{figure*}

\begin{figure}[t]
	\centering
    \begin{minipage}[t]{0.23\textwidth}
        \centering
        \includegraphics[trim=4.3cm 8.0cm 4.7cm 9cm, clip=true, scale=0.35]{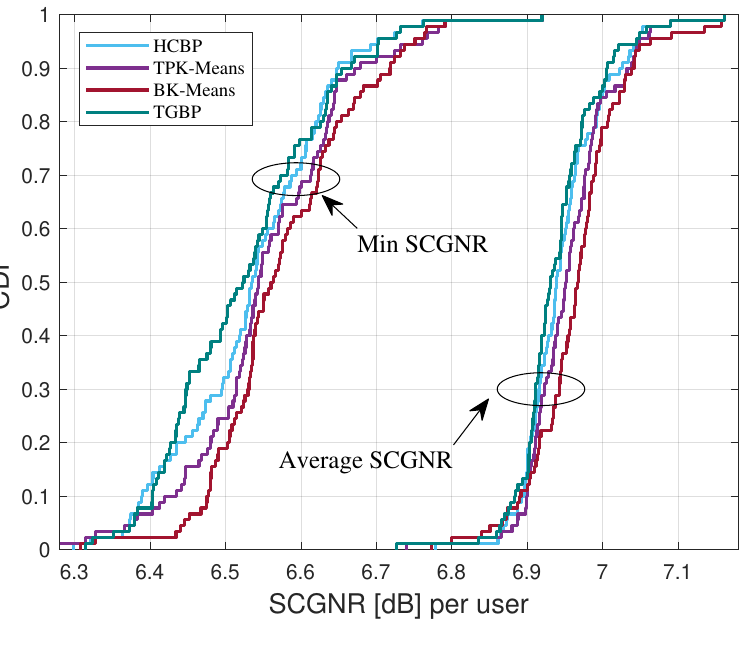}    	
        \subcaption{}
        \label{fig:SNR_small}
    \end{minipage}
    \hfill
    \begin{minipage}[t]{0.23\textwidth}
        \centering
        \includegraphics[trim=4.3cm 8.0cm 4.7cm 9cm, clip=true, scale=0.35]{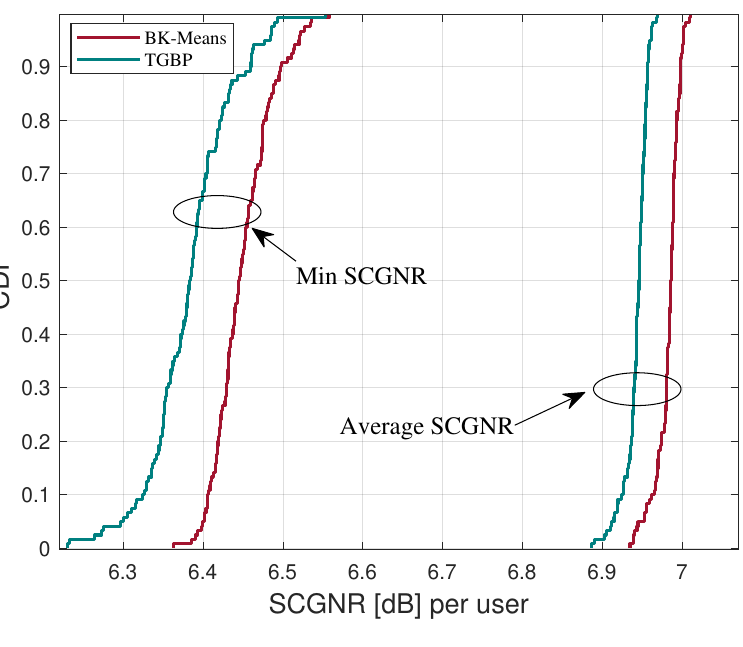}
        \subcaption{}
        \label{fig:SNR_large}
    \end{minipage}
\vspace{-0.2cm}
\caption{(a) CDF of the average SCGNR and min SCGNR [dB] per user on the small user set. (b) CDF of the average SCGNR and min SCGNR [dB] per user on the large user set.}
\label{fig:fig3}
\vspace{-0.5cm}
\end{figure}

\section{Clustering for Satellite Beam Placement}
\subsection{Problem Formulation}

We consider the beam placement problem in~\cite{10798457}, where a LEO satellite with multiple steerable beams serves a set of ground users. Each beam has a limited coverage area defined by the Half Power Beam Width (HPBW), and only a subset of beams can be simultaneously active.

Let $\mathcal{U} = \{1,\dots,N\}$ be the set of users, and $\mathcal{B} = \{1,\dots,M\}$ the candidate beam locations. A user $i$ can only be served by beam $j$ if it lies within its HPBW, indicated by $a_{ij} \in \{0,1\}$. Define $x_{ij} \in \{0,1\}$ as the assignment variable and $y_j \in \{0,1\}$ as the beam activation variable. The objective is to minimize the number of active beams:
\begin{equation}
\min \sum\nolimits_{j \in \mathcal{B}} y_j
\quad \text{s.t.} 
\sum\nolimits_{j \in \mathcal{B}} x_{ij} = 1,\quad
x_{ij} \leq a_{ij},
x_{ij} \leq y_j.
\end{equation}
This problem can be reformulated as a graph covering problem. Let $G = (V, E)$ be an undirected graph where each vertex in $V = \mathcal{U}$ corresponds to a user. Two users $i$ and $k$ are connected by an edge if their angular separation, as viewed from the satellite, satisfies:
\[
\alpha_{ik} = \arccos\left( \frac{2 S_i S_k}{S_i^2 + S_k^2 - d_{ik}^2} \right) \leq \frac{\alpha_{\max}}{2},
\]
where $S_i$, $S_k$ are the slant ranges and $d_{ik}$ is the geographical distance between users $i$ and $k$. The $\alpha_{\max}$ corresponds to the beam’s HPBW. The adjacency matrix $\mathbf{A} = [a_{ik}]$ is defined as: $a_{ik} =
1$ \text{if } $\alpha_{ik} \leq \alpha_{\max}/2,$ and is 
0 \text{otherwise}.

A group of users can be served by a single beam if and only if they form a clique in $G$, i.e., every pair is connected. Thus, minimizing the number of active beams is equivalent to covering $G$ using the minimum number of cliques, known as the \emph{Minimum Clique Cover (MCC)} problem. To address this NP-hard problem, the authors proposed two heuristics: BK-Means (clustering-based) and TGBP (greedy-based), evaluated against TPK-Means and HCBP.

\subsection{Numerical Results}

For simulations, we consider two scenarios: small user sets (10-30 users) and large user sets (100-1500 users). Note that TPK-Means and HCBP are only applicable to small user sets due to their high computational complexity. The maximum number of iterations is  $\mu=400$, and the inner maximum number of K-Means iterations in BK-Means is $I = 400$. Each beam center of the satellite has a maximum gain of 50 [dBi] and a HPBW with a maximum angle of $\alpha_{\text{max}} = 3.2^\circ$.

Fig.\ref{fig:fig2}(a) and Fig.\ref{fig:fig2}(b) show the number of active beams (NABs) for different algorithms, including HCBP, TPK-Means, BK-Means, and TGBP. The results indicate that TGBP generally requires fewer active beams, especially in large user sets. For small user sets, the performance of the algorithms is closer, but TGBP still has a slight advantage. Fig.\ref{fig:fig2}(c) and Fig.\ref{fig:fig2}(d)
display the load balancing gap, which is calculated as the difference between the beams serving the most and the least users. In the small user set, TGBP shows good load balancing. However, in the large user set, BK-Means achieves better load balancing with a smaller difference, suggesting a more even distribution of users across beams. Fig.\ref{fig:fig3}(a) and Fig.\ref{fig:fig3}(b) present the cumulative distribution function (CDF) of the statistical channel gain to noise ratio (SCGNR) for each user. BK-Means performs better in both average and minimum SCGNR compared to other algorithms in both small and large user sets. Compared to the baseline methods, both the modified K-Means and the graph-based approaches demonstrate strong potential for real-world applications. The modified K-Means improves SCGNR performance, while the graph-based method is more effective at minimizing the number of active beams.

\section{Challenges and Future Research}

Given the growing importance of clustering techniques in optimizing satellite communication system performance, further research is necessary to refine existing methods or develop novel approaches that can adapt to the dynamic and resource-constrained nature of satellite networks. To expand the practical applicability of clustering in future satellite systems, this section explores key research challenges in both user clustering and satellite clustering. A summary of the discussed challenges and potential research directions is also presented at the end of this section for clarity and future reference.

\subsection{Challenges}
\textit{$1$) For user clustering}: Satellite communication systems face major challenges due to uneven and time-varying user demand. Traffic often concentrates in specific areas or fluctuates over time, complicating resource allocation, especially in LEO constellations where satellites move rapidly and coverage constantly changes. While some studies~\cite{9443991, 9786763} adapt to traffic, they assume static or single-satellite scenarios, highlighting the need for dynamic models that capture satellite motion, user mobility, and shifting demand for efficient and fair service.


Another key challenge lies in adjusting beamwidth, which involves a trade-off between coverage and performance. 
Wider beams cover larger areas with fewer beams but suffer from interference and low efficiency, whereas narrower beams provide higher capacity and spatial reuse at the cost of greater management complexity and handover frequency~\cite{chen2024beam}. Dynamically selecting optimal beamwidth based on traffic and mobility patterns is essential, particularly in dynamic LEO constellations with constantly moving users and satellites.

\textit{$2)$ For satellite clustering}:  Signal synchronization is a significant challenge in satellite communications systems, especially in cooperative transmission. Due to the vast distances and varying propagation delays between satellites and ground terminals, signals from different satellites often arrive misaligned, leading to interference and reduced system performance \cite{chen2024asynchronous,humadi2024distributed}. Moreover, satellite clusters are inherently vulnerable to instability due to the unpredictable link failures, which can be caused by factors such as atmospheric interference, physical obstructions, or hardware degradation over time. Furthermore, the frequent changes in network topology, especially in LEO, make it challenging to maintain stable and continuous communication paths \cite{liu2021reliable}.
\vspace{-0.25cm}
\subsection{Future Research Directions}

\textit{$1)$ Game-Theoretic Clustering for Load Balancing and Fairness}: In satellite communication networks, especially user-centric clustering, users can be modeled as strategic players competing for limited satellite resources. When users selfishly select the best satellite for themselves, congestion, interference, or resource overload may occur. Future research should explore game-theoretic clustering frameworks (e.g., coalition games or congestion games) that incentivize cooperation or fair resource usage.

\textit{$2)$ Metaheuristic Clustering with Evolutionary Computation:}
Metaheuristic algorithms, particularly evolutionary computing methods like Genetic Algorithms (GA), offer promising solutions for clustering under highly dynamic conditions typical of LEO satellite networks. These methods can explore large search spaces efficiently and adapt to topology changes in near real-time while ensuring near optimal solution. Future directions may involve developing hybrid metaheuristic models, combining GA with swarm intelligence or deep learning to enable fast convergence and high-quality clustering, even under strict latency and power constraints.

\textit{$3)$ DBSCAN:} 
Recent studies have demonstrated that density-based clustering algorithms such as DBSCAN can effectively address user mobility and irregular cluster shapes in mmWave networks. Unlike traditional methods such as K-Means, DBSCAN does not require prior knowledge of the number of clusters and can identify outliers, making it particularly suitable for dynamic environments\cite{elsayed2020radio}.

\section{Conclusion}

This paper presented a comprehensive review and comparison of clustering techniques in satellite communication systems, analyzing both clustering criteria-focusing on user and satellite clustering-and a range of methods, including graph-based and machine learning approaches. We also discussed key challenges in satellite networks and highlighted emerging trends such as hybrid algorithms and real-time adaptive clustering. While this paper does not cover all aspects of the field, we hope it provides valuable insights for researchers and practitioners in understanding clustering strategies, guiding algorithm development, and exploring future opportunities in satellite communications.
\vspace{-0.2cm}
\section*{Acknowledgment}
This research is funded by
the Vietnam Ministry of Education and Training under project number B2025-BKA-04. 
\vspace{-0.25cm}

\bibliographystyle{IEEEtran}
\bibliography{IEEEabrv,refs}

\begin{thebibliography}{10}
\providecommand{\url}[1]{#1}
\csname url@samestyle\endcsname
\providecommand{\newblock}{\relax}
\providecommand{\bibinfo}[2]{#2}
\providecommand{\BIBentrySTDinterwordspacing}{\spaceskip=0pt\relax}
\providecommand{\BIBentryALTinterwordstretchfactor}{4}
\providecommand{\BIBentryALTinterwordspacing}{\spaceskip=\fontdimen2\font plus
\BIBentryALTinterwordstretchfactor\fontdimen3\font minus
  \fontdimen4\font\relax}
\providecommand{\BIBforeignlanguage}[2]{{%
\expandafter\ifx\csname l@#1\endcsname\relax
\typeout{** WARNING: IEEEtran.bst: No hyphenation pattern has been}%
\typeout{** loaded for the language `#1'. Using the pattern for}%
\typeout{** the default language instead.}%
\else
\language=\csname l@#1\endcsname
\fi
#2}}
\providecommand{\BIBdecl}{\relax}
\BIBdecl

\bibitem{guidotti2020architectures}
A.~Guidotti, S.~Cioni, G.~Colavolpe, M.~Conti, T.~Foggi, A.~Mengali,
  G.~Montorsi, A.~Piemontese, and A.~Vanelli-Coralli, ``{Architectures,
  standardisation, and procedures for 5G Satellite Communications: A survey},''
  \emph{Computer Networks}, vol. 183, p. 107588, 2020.

\bibitem{jung2024modeling}
D.-H. Jung and othersJ, ``{Modeling and analysis of GEO satellite networks},''
  \emph{IEEE Transactions on Wireless Communications}, 2024.

\bibitem{10507224}
D.~Kim \emph{et~al.}, ``{Coverage Analysis of Dynamic Coordinated Beamforming
  for LEO Satellite Downlink Networks},'' \emph{IEEE Transactions on Wireless
  Communications}, vol.~23, no.~9, pp. 12\,239--12\,255, 2024.

\bibitem{10713888}
H.~Deng \emph{et~al.}, ``{Satellites Beam Hopping Scheduling for Interference
  Avoidance},'' \emph{IEEE Journal on Selected Areas in Communications},
  vol.~42, no.~12, pp. 3647--3658, 2024.

\bibitem{park2025trends}
S.~Park, ``{Trends on the Expansion of the 6G Mobile Spectrum},''
  \emph{Electronics and Telecommunications Trends}, vol.~40, pp. 22--33, 2025.

\bibitem{song2023cluster}
Y.~Song, J.~Ou, J.~Wu, Y.~Wu, L.~Xing, and Y.~Chen, ``A cluster-based genetic
  optimization method for satellite range scheduling system,'' \emph{Swarm and
  Evolutionary Computation}, vol.~79, p. 101316, 2023.

\bibitem{10798457}
T.~V. Chien \emph{et~al.}, ``{Fast Beam Placement for Ultra-Dense LEO
  Networks},'' \emph{IEEE Wireless Communications Letters}, vol.~14, no.~3, pp.
  621--625, 2025.

\bibitem{xie2025multi}
X.~Xie, K.~Fan, W.~Deng, N.~Pappas, and Q.~Zhang, ``{Multi-Satellite Beam
  Hopping and Power Allocation Using Deep Reinforcement Learning},''
  \emph{arXiv preprint arXiv:2501.02309}, 2025.

\bibitem{10946453}
B.~Shen \emph{et~al.}, ``{Efficient and Secure Federated Dynamic Spectrum
  Access for LEO Satellite Internet-of-Things},'' in \emph{2024 IEEE 24th
  International Conference on Communication Technology (ICCT)}, 2024, pp.
  366--371.

\bibitem{bukhari2025k}
S.~M. Bukhari and W.-C. Song, ``{K-Means++ Clustering-Based Approach for SDN
  Controller Placement in LEO Satellite Networks},'' \emph{IEEE Access}, 2025.

\bibitem{jung2023satellite}
D.-H. Jung \emph{et~al.}, ``{Satellite clustering for non-terrestrial networks:
  Concept, architectures, and applications},'' \emph{IEEE Vehicular Technology
  Magazine}, vol.~18, no.~3, pp. 29--37, 2023.

\bibitem{liu2021reliable}
J.~Liu, X.~Zhang, R.~Zhang, T.~Huang, and F.~R. Yu, ``{Reliable and
  low-overhead clustering in LEO small satellite networks},'' \emph{IEEE
  Internet of Things Journal}, vol.~9, no.~16, pp. 14\,844--14\,856, 2021.

\bibitem{9295418}
A.~Ivanov, R.~Bychkov, and E.~Tcatcorin, ``{Spatial Resource Management in LEO
  Satellite},'' \emph{IEEE Transactions on Vehicular Technology}, vol.~69,
  no.~12, pp. 15\,623--15\,632, 2020.

\bibitem{10445284}
X.~Ding \emph{et~al.}, ``{Improving User Capacity of Satellite Internet of
  Things via Joint User Grouping and Multi-Beam Processing},'' \emph{IEEE
  Transactions on Communications}, vol.~72, no.~7, pp. 3957--3969, 2024.

\bibitem{10648635}
B.~Liu, L.~Kuang, and J.~Lu, ``{Earth-Fixed Multicast User Subgrouping for NGSO
  Satellite With Phased Array Antenna},'' \emph{IEEE Internet of Things
  Journal}, vol.~11, no.~23, pp. 37\,661--37\,674, 2024.

\bibitem{guidotti2020clustering}
A.~Guidotti and A.~Vanelli-Coralli, ``{Clustering strategies for multicast
  precoding in multibeam satellite systems},'' \emph{International Journal of
  Satellite Communications and Networking}, vol.~38, pp. 85--104, 2020.

\bibitem{10034945}
Y.~Li, S.~Zhu, and J.~Dai, ``{Joint User Grouping and Resource Allocation for
  LEO Satellite Multicast},'' \emph{IEEE Systems Journal}, vol.~17, no.~3, pp.
  4695--4702, 2023.

\bibitem{10433460}
K.~Wang, B.~Feng, J.~Zhao, W.~Lin, Z.~Deng, D.~Wang, Y.~Cen, and G.~Wu,
  ``Channel correlation based user grouping algorithm for nonlinear precoding
  satellite communication system,'' \emph{China Communications}, vol.~21,
  no.~1, pp. 200--214, 2024.

\bibitem{10601666}
S.~Kashyap and N.~Gupta, ``{Demand-Based Dynamic Bandwidth Allocation in
  Multi-Beam Satellites Using Machine Learning Concepts},'' \emph{Intelligent
  and Converged Networks}, vol.~5, no.~2, pp. 147--166, 2024.

\bibitem{10460317}
N.~Pachler, E.~F. Crawley, and B.~G. Cameron, ``{A Scalable Algorithm for User
  Grouping in LEO-MEO-HEO Hybrid Very High Throughput Satellite
  Constellations},'' \emph{IEEE Transactions on Cognitive Communications and
  Networking}, vol.~10, no.~4, pp. 1511--1524, 2024.

\bibitem{zhu2022service}
X.~Zhu and Z.~Wang, ``{Service Throughput Oriented Clustering in LEO Satellite
  Networks},'' in \emph{2022 4th International Conference on Frontiers
  Technology of Information and Computer (ICFTIC)}.\hskip 1em plus 0.5em minus
  0.4em\relax IEEE, 2022, pp. 842--850.

\bibitem{10648888}
D.~Wang, H.~Qin, Y.~Zhang, Y.~Yang, and H.~Lv, ``{Fast Clustering Satellite
  Selection Based on Doppler Positioning GDOP Lower Bound for LEO
  Constellation},'' \emph{IEEE Transactions on Aerospace and Electronic
  Systems}, vol.~60, no.~6, pp. 9401--9410, 2024.

\bibitem{jing2025fast}
D.~Jing \emph{et~al.}, ``{A Fast Satellite Selection Algorithm Based on
  Hierarchical Clustering and Iterative Subset Optimization},'' \emph{Remote
  Sensing}, vol.~17, no.~5, p. 853, 2025.

\bibitem{geng2021resilient}
S.~Geng, S.~Liu, and Z.~Fang, ``Resilient communication model for satellite
  networks using clustering technique,'' \emph{Reliability Engineering \&
  System Safety}, vol. 215, p. 107850, 2021.

\bibitem{tung2024graph}
N.~X. Tung \emph{et~al.}, ``{Graph Neural Networks for Next-Generation-IoT:
  Recent Advances and Open Challenges},'' \emph{arXiv preprint
  arXiv:2412.20634}, 2024.

\bibitem{10018884}
S.~Li, D.~Ye, Z.~Sun, J.~Zhang, and W.~Zhong, ``{Collision-Free Flocking
  Control for Satellite Cluster With Constrained Spanning Tree Topology and
  Communication Delay},'' \emph{IEEE Transactions on Aerospace and Electronic
  Systems}, vol.~59, no.~4, pp. 4134--4146, 2023.

\bibitem{9914723}
D.~G. Riviello, B.~Ahmad, A.~Guidotti, and A.~Vanelli-Coralli, ``{Joint
  Graph-based User Scheduling and Beamforming in LEO-MIMO Satellite
  Communication Systems},'' in \emph{2022 11th Advanced Satellite Multimedia
  Systems Conference and the 17th Signal Processing for Space Communications
  Workshop (ASMS/SPSC)}, 2022, pp. 1--8.

\bibitem{5453745}
S.~Na, L.~Xumin, and G.~Yong, ``{Research on k-means Clustering Algorithm: An
  Improved k-means Clustering Algorithm},'' in \emph{2010 Third International
  Symposium on Intelligent Information Technology and Security Informatics},
  2010, pp. 63--67.

\bibitem{10133823}
T.~V. Chien \emph{et~al.}, ``{Phase Shift Design for RIS-Aided Cell-Free
  Massive MIMO With Improved Differential Evolution},'' \emph{IEEE Wireless
  Communications Letters}, vol.~12, no.~9, pp. 1499--1503, 2023.

\bibitem{10794213}
Y.~He, K.~Qi, X.~Feng, and R.~Chen, ``{Load-Balanced and Full-Coverage Beam
  Layout for Multibeam Satellite Systems},'' in \emph{2024 6th International
  Conference on Communications, Signal Processing, and their Applications
  (ICCSPA)}, 2024, pp. 1--5.

\bibitem{camino2021milp}
J.-T. Camino, C.~Artigues, L.~Houssin, and S.~Mourgues, ``{MILP formulation
  improvement with k-means clustering for the beam layout optimization in
  multibeam satellite systems},'' \emph{Computers \& Industrial Engineering},
  vol. 158, p. 107228, 2021.

\bibitem{chen2023edge}
C.-Y. Chen \emph{et~al.}, ``{Edge selection and clustering for federated
  learning in optical inter-leo satellite constellation},'' in \emph{2023 IEEE
  34th Annual International Symposium on Personal, Indoor and Mobile Radio
  Communications (PIMRC)}.\hskip 1em plus 0.5em minus 0.4em\relax IEEE, 2023,
  pp. 1--6.

\bibitem{9443991}
L.~Wang \emph{et~al.}, ``{Dynamic Beam Hopping of Multi-beam Satellite Based on
  Genetic Algorithm},'' in \emph{2020 IEEE Intl Conf on Parallel \& Distributed
  Processing with Applications, Big Data \& Cloud Computing, Sustainable
  Computing \& Communications, Social Computing \& Networking
  (ISPA/BDCloud/SocialCom/SustainCom)}, 2020, pp. 1364--1370.

\bibitem{9786763}
T.~Ramírez, C.~Mosquera, and N.~Alagha, ``{Flexible User Mapping for Radio
  Resource Assignment in Advanced Satellite Payloads},'' \emph{IEEE
  Transactions on Broadcasting}, vol.~68, no.~3, pp. 723--739, 2022.

\bibitem{chen2024beam}
J.~Chen, Q.~Jiang, and M.~Yan, ``{A Beam Hopping Scheme Based on Adaptive Beam
  Radius for LEO Satellites},'' \emph{Sensors (Basel, Switzerland)}, vol.~24,
  no.~20, p. 6574, 2024.

\bibitem{chen2024asynchronous}
X.~Chen and Z.~Luo, ``{Asynchronous Interference Mitigation for LEO
  Multi-Satellite Cooperative Systems},'' \emph{IEEE Transactions on Wireless
  Communications}, 2024.

\bibitem{humadi2024distributed}
K.~Humadi \emph{et~al.}, ``{Distributed Massive MIMO System with Dynamic
  Clustering in LEO Satellite Networks},'' in \emph{2024 6th International
  Conference on Communications, Signal Processing, and their Applications
  (ICCSPA)}.\hskip 1em plus 0.5em minus 0.4em\relax IEEE, 2024, pp. 1--6.

\bibitem{elsayed2020radio}
M.~Elsayed and M.~Erol-Kantarci, ``Radio resource and beam management in 5g
  {mmWave} using clustering and deep reinforcement learning,'' in
  \emph{GLOBECOM 2020-2020 IEEE Global Communications Conference}.\hskip 1em
  plus 0.5em minus 0.4em\relax IEEE, 2020, pp. 1--6.

\end{thebibliography}
\end{document}